# Experimental Study of Neutron Formation in X-Ray Discharge: Is There a Challenge to Standard Model?

Max I. Fomitchev-Zamilov[1], Ross Koningstein[2]

[1]Maximus Energy Corporation, [2]Google Corporation

Email: founder@maximus.energy

**ABSTRACT**

In this paper we examine the puzzle of neutron formation in atmospheric discharge. While conventional explanation relies on photonuclear reactions, voltages used to recreate such discharges in the lab (1 MV) are thought to be not high enough to accelerate electrons beyond the threshold of photonuclear reactions (10.5 MeV).

We have proposed a solution to the neutron formation puzzle and point out that rapidly changing magnetic vector potential creates strong electrokinetic field that can accelerate free electrons to energies that are much higher than what is possible with the voltage of the discharge.

We report on the development of highly sophisticated and extremely sensitive neutron detector system that can reliably detect neutron fluxes smaller than 0.1 CPS above background and is completely impervious to EM noise.

Finally, we present measurements of weak yet statistically significant neutron fluxes originating from sustained 20-30 kV / 20-30 mA X-ray producing discharge in dilute atmosphere.

**INTRODUCTION**

According to the Standard Model of particle physics probability of neutron formation due to $p^+ + e^- \rightarrow n^0$ reaction is vanishingly small [1]. This is not surprising since this reaction is mediated by weak force and requires an anti-neutrino that must first interact with an electron to form a $W^-$ boson ($e^- + \nu^0 \rightarrow W^-$), which then can interact with a proton to form a neutron ($p^+ + W^- \rightarrow n^0$). But neutrinos are called neutrinos for a reason - due to their miniscule reaction cross sections (on the order of $10^{-44}$ cm$^2$ [2]) probabilities of their interactions with matter are virtually zero.

Inverse beta decay of a proton ($p^+ + \nu^0 \rightarrow e^+ + n^0$) can also occur and in is in fact commonly utilized for neutrino detection [3]. However, the likelihood of this interaction is also very low. To derive a detectable signal, hundreds of cubic meters of liquid scintillator are required.

One thing appears to be clear: neutrino oscillations[ref?] indicate non-zero neutrino masses whereas the Standard Model requires neutrinos to be massless. There have been multiple proposals to extend the Standard Model [4]. It is important to keep looking for and keep accumulating experimental evidence [5, 6] that one day will help supercede the currently accepted Standard Model with a more general theory.

In this regard we turned our attention to an experiment that on one hand may challenge the Standard Model and on another hand does not require a multi-million dollar infrastructure typically associated with particle physics. Our curiosity was peaked by multiple reports describing elevated levels of neutron counts associated with thunderstorms [2, 3]. To make matters even more intriguing positrons have been detected in thunderstorm clouds [7] and they appear to be associated with terrestrial gamma flashes [8]. This neutron, positron and gamma radiation clearly indicates that nuclear processes occur naturally and are driven by weather.

The recently published report by Enoto et al. [9] greatly improved our understanding of thunderstorm-related nuclear processes by painting a convincing picture where neutrons originate due to photonuclear spallation of nitrogen: $^{14}N + \gamma \rightarrow {}^{13}N + n^0$. Nitrogen-13 further decays into carbon-13 via positron emission: $^{13}N \rightarrow {}^{13}C + e^+$ whereas the neutrons are typically captured by nitrogen nuclei that either turn to carbon-14: $^{14}N + n^0 \rightarrow {}^{14}C + p^+$ (96%) or de-excite by emitting a gamma ray: $^{14}N + n^0 \rightarrow {}^{15}N + \gamma$ (4%). The conclusions of the Japanese team are well supported by the gamma spectra that they recorded during numerous thunderstorm events: shape temporal behavior of these spectra are fully consistent with the aforementioned reactions.

In the same time Russian team lead by Agafonov decided to study an artificial lightning in the lab by passing a 1 megavolt / 10 kiloamp discharge between aluminum electrodes in air [10]. The Russian team also detected neutrons correlating with X-rays (there were no neutrons without X-rays but sometimes there were X-rays without neutrons). However, after the detailed theoretical analysis they concluded that there is no known mechanism that could explain their observations within the framework of the Standard Model [11]. This conclusion is not surprising since 1 megavolt potential is simply not large enough to accelerate electrons and ions to energies that could give rise to neutron-forming nuclear reactions (i.e. the threshold energy for photonuclear disintegration of nitrogen is 10.5 MeV).

Motivated by these findings we decided to conduct experiments as follows:

1) Study continuous high-voltage (20-50 kV) / low current (10-40 mA) X-ray producing discharges in dilute gases;

2) Study pulsed high-voltage (30-100 kV) / high-current (10-30 kA) discharges at atmospheric pressure.

In both sets of the experiments we have put our emphasis on developing tools and techniques that would allow for unprecedented sensitivity in detecting extremely weak neutrons fluxes. The expectation was that by increasing neutron detection sensitivity by a factor of $10^4$ compared to Agafonov et al. [10] and combining it with vastly extended (by a factor of $10^{10}$) counting time we should be able to detect and quantify even the lowest probability nuclear processes.

In this paper we report the results of the first experimental group while the work on the second group of experiments is still ongoing.

**EXPERIMENTAL SETUP**

The general idea behind the experiment design was as follows: the work of Agafonov et al. indicated that neutrons are generated during dark phase of atmospheric discharge when the voltage is still high and the discharge current has not yet reached peak [10]. This phase of discharge lasts for about 100 nanoseconds as voltage drops rapidly with the increase in current. Also, this phase is associated with X-ray emission, and the Agafonov et al. team detected neutrons only when X-rays were present; neutrons were never detected when X-rays failed to register. In attempt to increase duration of the X-ray generating phase of discharge we adopted a gas-filled cold-cathode X-ray tube design to model the 'dark phase' of the discharge - Fig. 1. The key advantage of the X-ray tube design is that it allows sustaining discharge almost indefinitely while providing means for precise control of voltage and current by adjusting pressure within the tube. The removable cathode allowed exploring different materials and determine if surface composition played a significant role in the purported nuclear effect. With this apparatus, where the X-ray tube is not immerse t oil the max voltage was limited by length of the ceramic break separating negative cathode potential from the grounded anode. On air the tube would begin to arc along the ceramic break's surface at 50 kV.

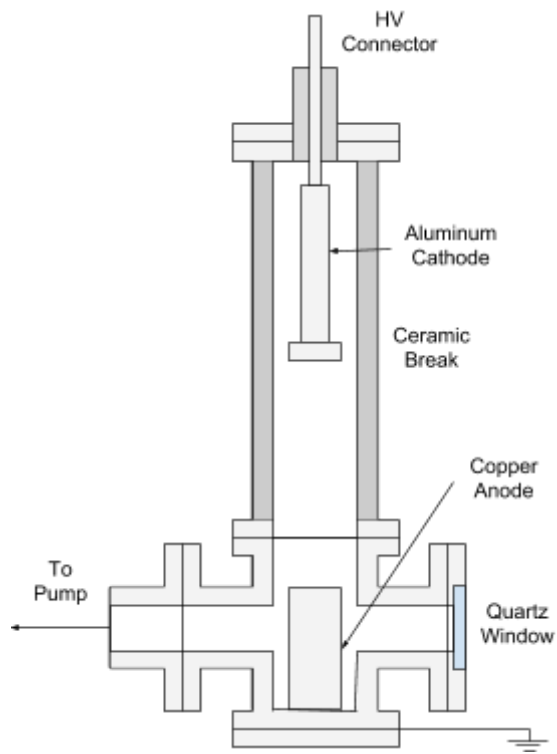

Fig. 1. Cold-cathode X-ray tube used in our experiments.

Fig. 2 shows our experimental setup. We attached the x-ray tube to a gas cylinder (typically hydrogen) via leak valve L1. To control pressure we further attached the tube via leak valve L2 to a vacuum system comprising a turbomolecular pump (Varian Turbo-V70LP), full-range vacuum pressure gauge (Pfeiffer PKR251), and MKS HPQ2 residual gas analyzer (RGA).

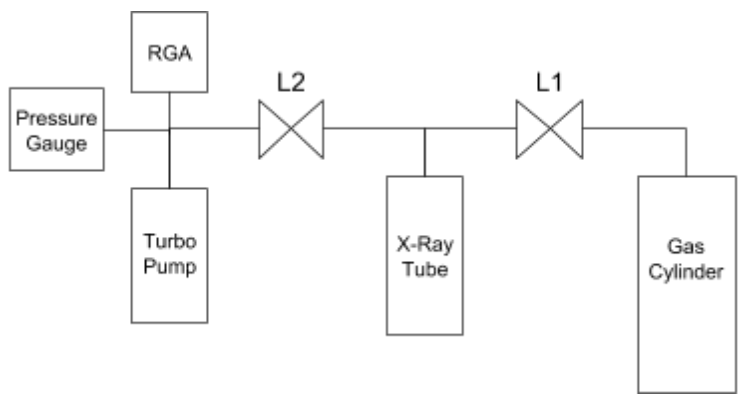

Fig. 1. Block diagram of the experimental setup.

We have experimented with two different modes of powering of the X-ray tube: AC and DC. In the DC mode (Fig. 3a) we used Spellman X-ray regulated power supplies (XLF, SL or DXM series) connected directly to the tube. These power supplies automatically maintained a desired negative voltage between cathode and anode while allowing us to set current by adjusting pressure inside the tube by operating the leak valve L1 (see Fig. 2). The high voltage DC power supplies we employed were rated at 600-1200 watts and allowed controlling voltage in the range of 0..50 kV and current in the range of 0..40 mA.

In the AC mode we connected the X-ray tube directly to the output of high voltage transformer with primary controlled by a variac (Fig. 3b). Although the transformer was capable of generating peak AC voltages up to 120 kV we typically worked with RMS voltages in the range of 20-35 kV and RMS currents in the range of 10-30 mA.

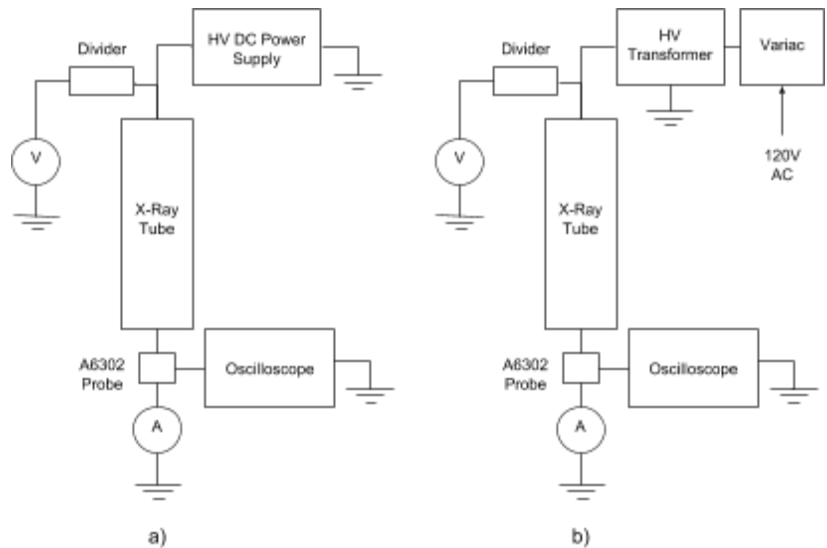

Fig. 3. a) DC and b) AC mode of powering of the X-ray tube.

In all experiments we measured cathode voltage using 100:1 resistive divider and obtained RMS current readings via analog AC/DC ammeter connected in series with the anode of the X-ray tube. To capture tube current traces we used Tektronix A6302 current probe installed around the anode grounding wire.

It is noteworthy that in the AC mode the tube was self-rectifying and thus allowed current only in one direction (i.e. electrons could travel only from cathode to anode and not the other way around). Typical voltage and current traces for the AC mode operation are given on Fig. 4.

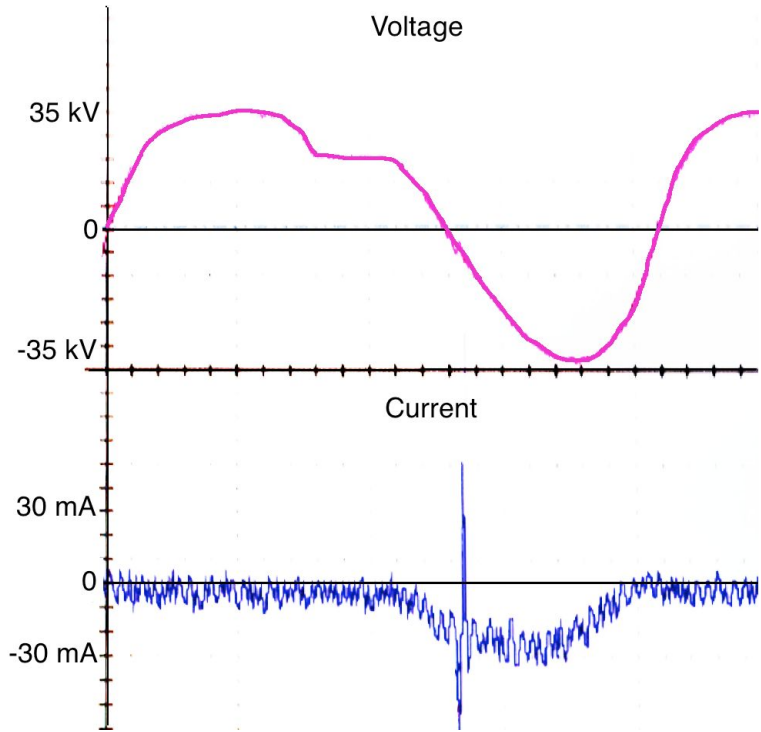

Fig. 4. Typical voltage and current trances during the AC mode operation of the X-ray tube.

Thus by adjusting the DC power supply or by turning the variac we could establish the cathode potential in the range of 0 to 50 kV RMS and by finely adjusting the pressure inside the tube (typically ~0.1 Torr) we could control the current in the range of 0 to 30 mA RMS.

The main difference between our setup and that of Agafonov et al. [10] is that we operated the discharge continuously but at much lower discharge voltage and current (20-30 kV vs 1 MV and 10-20 mA vs 10 kA). The expectation was that by increasing the neutron detection sensitivity by a factor of $10^4$ and by increasing duration of the experiment from 100 ns to 40 minutes (i.e. by a factor of $10^{10}$) we would still be able to detect nuclear events.

As a first-order approximation the probability of all nuclear reactions is an exponential function on energy. However, the unknown nature of the reactions in question did not allow us evaluating the exponential factor to determine how the reaction rate might change with the decrease in voltage and current. Therefore there was no way of knowing if the 14 orders of magnitude improvement arising from the increase in neutron detection sensitivity and measurement time would be adequate to compensate for 1.5 orders of magnitude reduction in voltage and 6 orders of magnitude reduction in current. Still, the general contrast of magnitudes (14 vs 7.5) provided a reason for optimism and therefore we proceeded with the experiment.

**THEORETICAL CONSIDERATIONS**

As we pointed out already, the observations of the Japanese team [9] are fully consistent with the photonuclear theory of neutron formation within the framework of the Standard Model. However, the observations of the Russian team are not [11]. Although from basic energy considerations electrons accelerated to 1 million volt potential possess energies in the nuclear range (1 MeV), the hypothesized photonuclear reactions require energies that are at least 10 times larger (e.g. >10.5 MeV). While the potential difference of 10 megavolts is plausible for lightning, the experiment conducted by the Russian team had a potential drop of only 1 million volts, hence the puzzle.

We submit that vector potential of magnetic field of the discharge holds the key to solving this puzzle. When the discharge current is changing (e.g. during onset of the discharge, due to current oscillations during the discharge, or during the final stage of the discharge) changing vector potential of the discharge's massive magnetic field will create a strong transient electric field given by:

$$E = - \partial A / \partial t \qquad (1)$$

For the sake of clarity we shall follow Jefimenko [13] and adopt the term 'electrokinetic' when referring to this field not to be confused with the electric field causing the discharge. This electrokinetic field will accelerate free electrons and ions the vicinity of the discharge and impart onto them significant additional momentum given by [13]:

$$\Delta P = -q \Delta A \qquad (2)$$

Thus, when the discharge current is rising positive ions will accelerate in the direction opposite to that of the discharge. When the discharge current is falling negative ions and free electrons will accelerate in the direction of the discharge. This behavior is nothing new and corresponds to a textbook illustration of a well-known mechanism of electromagnetic induction. What is new here is the realization that given large currents the momentum change due to the electrokinetic field can be very large and thus should not be ignored.

While the detailed theoretical treatment of the electrokinetic field is outside the scope of this paper we propose the following 'back of the envelope' calculation to illustrate the magnitude of the effect. I.e. neglecting retardation, for a discharge current of length $L$ and for an electron located at distance $R$ from the axis of the discharge the net change in momentum due to change in current $\Delta I$ is [13]:

$$|\Delta P| = e \, \Delta I \, \mu / 2\pi \, \ln(L/2R) \qquad (3)$$

Assuming that the electron was initially at rest the corresponding gain in kinetic energy is:

$$\Delta E = \Delta P^2/2m_e = (e\,\Delta I\,\mu/2\pi\,\ln(L/2R))^2/2m_e \quad (4)$$

Evaluating the equation (4) for a $\Delta I$ = 1000 A, $L$ = 1 meter, $R$ = 1 micron we obtain $\Delta E$ = 700 keV. Setting $\Delta I$ = 3000 A (i.e. 10% of the discharge current of 30 kA) results in $\Delta E$ = 6.5 MeV.

Clearly, relativistic effects, retardation and finite propagation time of fields must be carefully considered when evaluating general solution to Maxwell equations (also known as Jefimenko equations [18]):

$$E = 1/4\pi\varepsilon_0 \int \{[\varrho/r^3] + 1/r^2 c\,[\partial\varrho/\partial t]\}r\,dV - 1/4\pi\varepsilon_0\,c^2 \int 1/r\,[\partial J/\partial t]\}r\,dV \quad (5)$$

Where square brackets indicate that the quantities within the brackets are to be evaluated at a retarded time $t' = t - r/c$.

For now our objective was to show that a fairly small (10%) change in magnitude of discharge current can accelerate free electrons to energies far in excess of those associated with the original discharge potential and perhaps over the threshold of photonuclear reactions.

Also, from equation (3) it follows that in order to maximize neutron yield one has to have a powerful, narrowly focused yet highly unstable discharge.

**NEUTRON DETECTION**

Because our objective was to detect a very weak neutron signal we had to rule out such common means of neutron detection as CR-39 and BubbleTech BD-PND / BDT bubble detectors and opt for high-pressure Helium-3 proportional counter tubes due to their exceptionally high sensitivity to thermal neutrons. To maximize the solid angle coverage we used initially one and later three banks of 10x SNM-18 neutron detectors (12" long, 1.25" diameter, 4 bars Helium-3 pressure).

The neutron detector bank block diagram is shown on Fig. 5 and it contained the following components:

1) Block of 10x neutron detectors fitted with high-voltage input SHV connector to provide DC bias and and two BNC output connectors to provide a low-voltage, high-pass filtered output signals from each sub-block of 5x SNM-18 proportional counter tubes - Fig. 6;

2) HV Power Supply (Analog Technologies AHV5V3KV1MAW) that provided 1600V DC bias for the neutron detector tubes;

3) The bank produced detector output in the form of a negative pulse in the range from 0 to -200 mV (pulse duration 80 µs, rise time ~5 µs - FIg. 7) that was continuously digitized on two channels by PicoScope PS2204A DAQ at 1 MHz / 8 bit;

4) The data acquisition and signal analysis process was controlled by Insignia Tablet PC running PulseCounter software that we developed; the software performed data acquisition, detector signal discrimination and real-time statistical analysis of the acquired data.

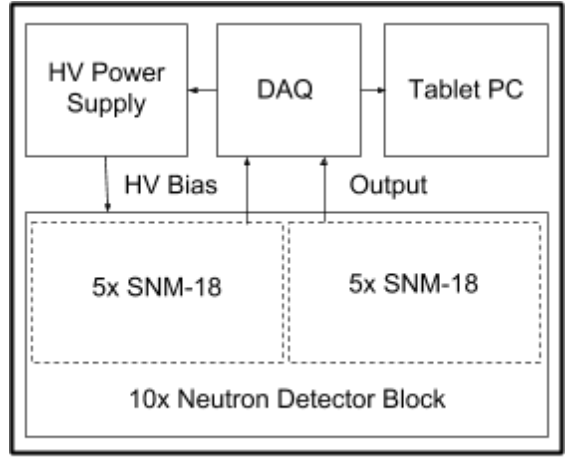

Fig. 5. Neutron detector bank.

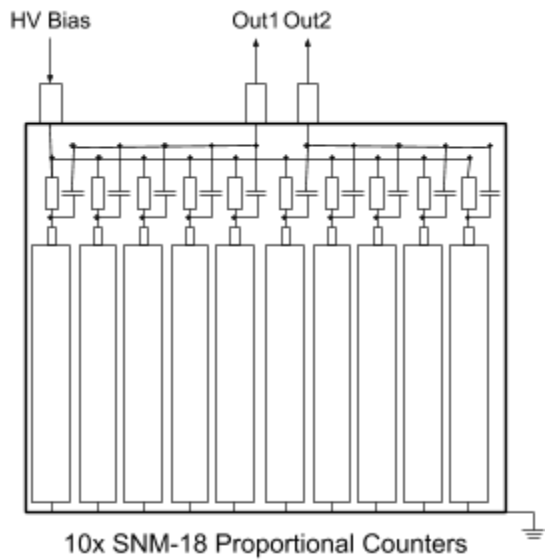

FIg. 6. Proportional counter bank diagram.

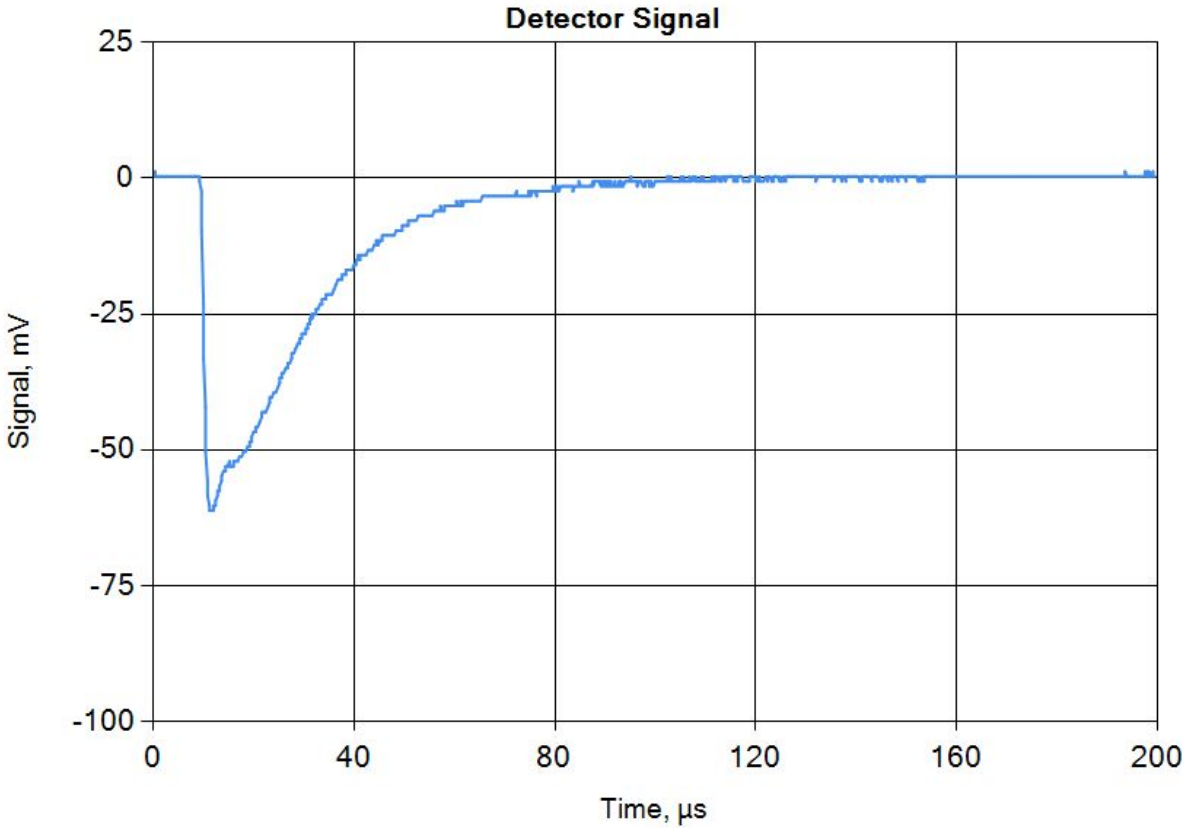

Fig. 7. SNM-18 proportional counter signal.

To eliminate electromagnetic interference we have enclosed each bank in a 3-mm thick aluminum Faraday enclosure (during earlier experiments we have also used 3-mm welded steel box). The resulting setup was completely impervious to electromagnetic noise: our data acquisition (DAQ) circuit registered less than 1 mV RMS of noise during test runs.

To ensure maximum detection probability we have located the three banks as close to the X-ray tube as we could without touching the high-voltage components - Fig. 8. To moderate any high-energy neutrons originating from the experiment we have installed 2.5" thick HDPE blocks between the X-ray tube and the banks. To screen X-rays we installed a ¼" lead sheet on the X-ray tube facing surface of the neutron detector bank.

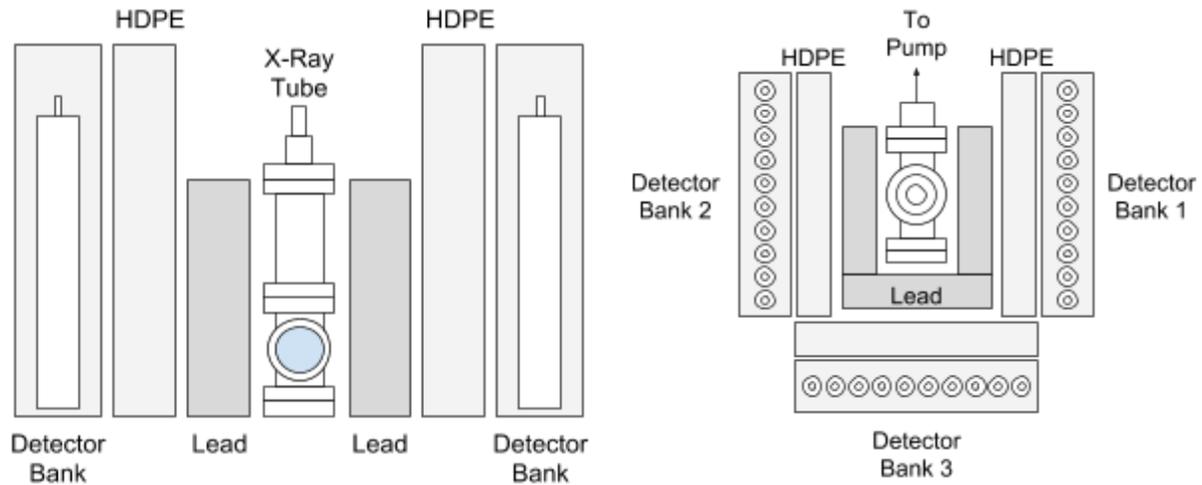

Fig. 8. Neutron detection setup: Left - sectional view; Right - top view.

The PulseCounter neutron pulse discrimination / counting algorithm was as follows:

- The level of the continuously acquired DAQ signal was monitored and when it exceeded the specified trigger level (typically -15 mV) the software increased the pulse count and performed pulse-shape integration to estimate energy;

- Pulses that failed to satisfy minimum (5 µs) or maximum (200 µs) width requirement were rejected and excluded from the total count;

- Each accepted pulse was stored in a PNG file for pulse-shape auditing purposes;

- Software constructed a 1000-channel uncalibrated energy spectrum of the acquired pulses; this was necessary to discriminate against x-rays.

It is common knowledge that helium-3 proportional counter tubes respond to X-rays / gammas as well as to neutrons. X-ray events typically form a gamma peak on the tube's energy spectrum - Fig. 9.

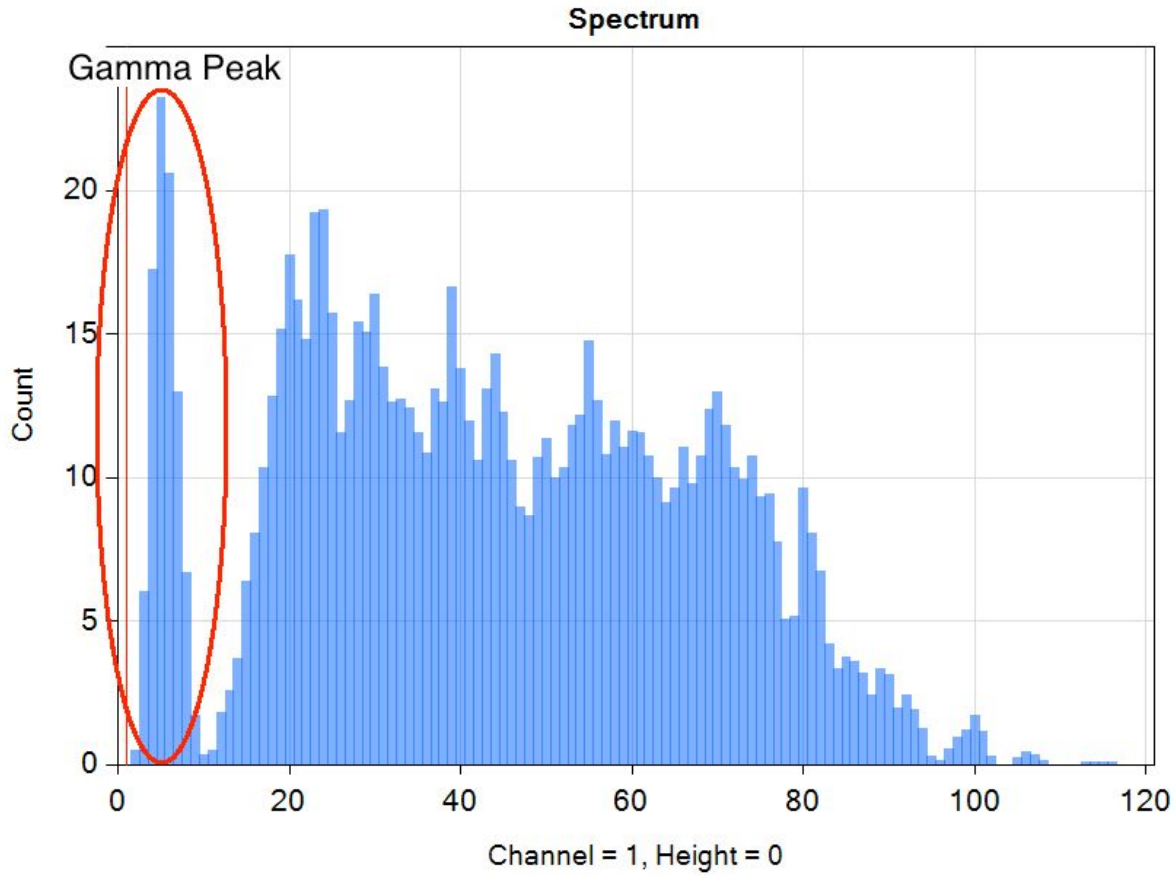

Fig. 9. Gamma peak on neutron spectrum.

Typical gamma rejection approach is to increase the discriminator level (level trigger in our case) until the x-ray peak is completely eliminated from the spectrum. Thus we choose to ignore low-energy pulses and assume that all such pulses are x-rays. There are two obvious problems with this approach: we unavoidably discard the lowest-energy neutron events along with X-rays and we cannot completely eliminate all gammas as high-energy (or high-intensity!) events will be indistinguishable from neutrons. This is a very-well known problem in proportional neutron counting that has no perfect solution. The only two practical approaches are: 1) screen proportional counter tubes with lead of thickness sufficient to eliminate the interfering X-rays, and 2) to monitor gamma spectrum during the experiment in order to ascertain gamma energy and gamma flux intensity and adjust the gamma rejection level accordingly. We used both of these approaches.

Finally, in order to establish statistical significance of neutron counts we captured background and experiment counts in interleaved sequences and recorded both CPS and CPM counts to obtain sufficiently large (>20 data points) sample sets for P-value calculation.

Because we have been looking for a very weak signal against a strongly variable background the size of the sample was critical. In the presence of the genuine neutron signal our protocol would yield a continuous improvement in P-value with the increase of the sample size: i.e. the more background and experiment sequences we capture the lower the P-value we will get assuming that there was indeed an excess of counts associated with the experimental sequences (regardless of how small this excess was). In practice, acquiring ten 10-minute sequences (5 background and 5 experiment) allowed us to unambiguously (better than 5% significance) identify neutron fluxes on the order of 0.1 CPS above background, which is an unprecedented level of sensitivity for any neutron detection system. In fact this real-time statistical analysis have allowed us to discover and eliminate a systematic error resulting from small (~ 1mV) periodic noise in our data acquisition system (Fig. 10) that boosted counts every other minute by pushing some would-be-rejected pulses over the trigger threshold (e.g. if a 15-mV pulse is normally rejected the periodic noise will make it look like 16-mV signal and thus bypass the rejection threshold).

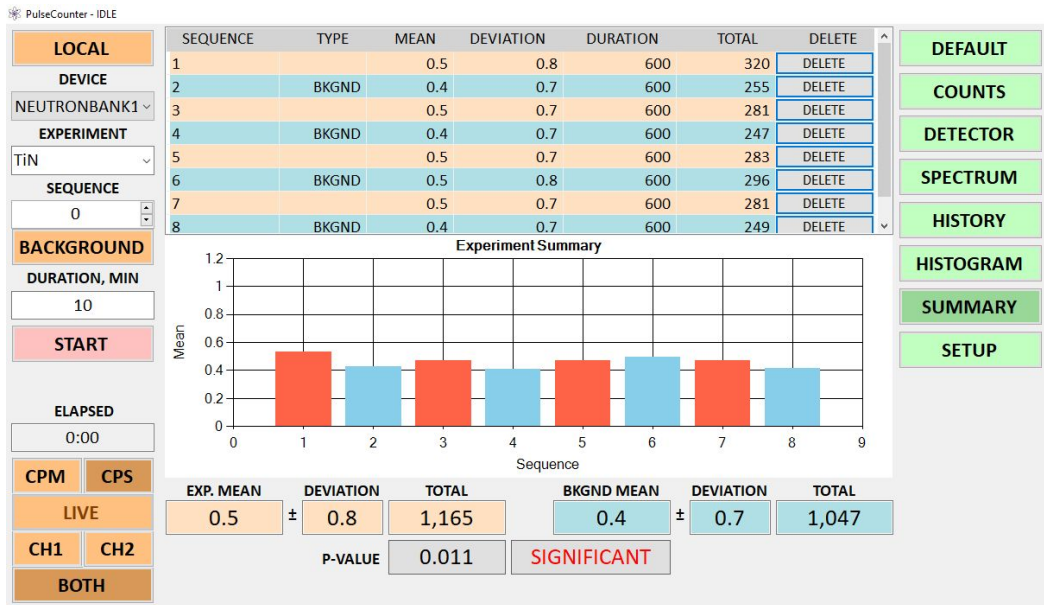

Fig. 10. False neutron counts due to very small periodic noise: 'experiment' counts appear larger than 'background' counts by 0.1 CPS at 1.1% significance level.

We have discovered this systematic by running an experiment and interleaving the background and the experiment counting sequences without actually conducting the experiment, i.e. we simply labeled every other sequence as 'experiment' and computed statistical significance, which due to indistinguishable nature of the 'background' and the 'experiment' labeled counts should have been around 50%. The significant departure of the P-value from 50% that decreased with the increase of the sample size revealed presence of small periodic noise that affected the counts.

**EXPERIMENTAL RESULTS**

Throughout 2017-2018 we have conducted numerous experiments in which we passed a 20-30 kV / 20-30 mA discharge through the cold-cathode X-ray tube and measured neutron counts when the discharge was on ('experiment') and when the discharge was off ('background'). A typical experiment contained 5-10 'experiment' and 5-10 'background' interleaved sequences with each sequence lasting for 10 minutes. During each experiment we tried maintaining voltage and current steady (the goal was 30 kV / 20 mA) although this was not always possible. In vast majority of the experiments we filled the tube with hydrogen and employed copper anode. Virtually all such experiments yielded no statistically-significant above-background neutron counts. Undeterred we decided to try different anode materials and on several occasions substituted hydrogen for water vapor. Some of these experiments yielded above-background neutron counts. All positive and select null experiments are summarized in Table 1.

| Date | Anode | Voltage, kV | Current, mA | Mode | Gas | Counts, CPS | Bkgnd, CPS | P-value |
|---|---|---|---|---|---|---|---|---|
| 5/9/2018 | Nickel 20Cb3 | 20 | 20 | AC | $H_2$ | 1.20 | 1.27 | n/a |
| **5/1/2018** | **AlN** | **0** | **0** | **n/a** | $H_2$ | **0.50** | **0.48** | **0.024** |
| **4/3/2018** | **AlN** | **25** | | **AC** | $H_2$ | **0.60** | **1.20** | |
| **4/28/2017** | **Nickel 20Cb3** | **~30** | **~15** | **AC** | $H_2$ | **0.65** | **0.48** | **0.007** |
| **4/27/2017** | **Stainless** | **~30** | **~15** | **AC** | $H_2$ | **0.69** | **0.58** | **0.0139** |
| **4/26/2017** | **Titanium** | **~30** | **~15** | **AC** | $H_2$ | **0.78** | **0.58** | **0.003** |
| 8/14/2017 | Copper | ~30 | ~15 | AC | $H_2$ | 0.28 | 0.29 | 0.47 |
| 8/14/2017 | Silica on Nickel | 25-29 | 10-15 | AC | $H_2$ | 0.28 | 0.28 | 0.47 |
| 8/7/2017 | Copper | 20 | 10-20 | AC | $H_2O$ | 0.35 | 0.35 | 0.48 |
| **8/6/2017** | **Titanium** | **~30** | **15** | **AC** | $H_2O$ | **0.39** | **0.35** | **0.0009** |
| **4/28/2017** | **Nickel 20Cb3** | **30** | **20** | **DC** | $H_2$ | **0.65** | **0.48** | **0.0007** |
| **4/26/2017** | **Titanium** | **15** | **1** | **DC** | $H_2$ | **0.78** | **0.58** | **0.0003** |

| 4/23/2017 | Stainless | 10-15 | 20-30 | DC | $H_2$ | 0.69 | 0.58 | 0.013 |

Table 1. Summary of experiments.

Typical voltage and current waveforms for AC excitation are shown on Fig. 4. Typical residual gas mass spectrum is shown on Fig. 11. Typical x-ray and gamma spectra (captured using Amptek XR-100SDD and Scionix 38B57 NaI(Tl) detectors respectively) are shown on Fig. 12 and 13. We did not spot any anomalies in RGA, x-ray or gamma spectra.

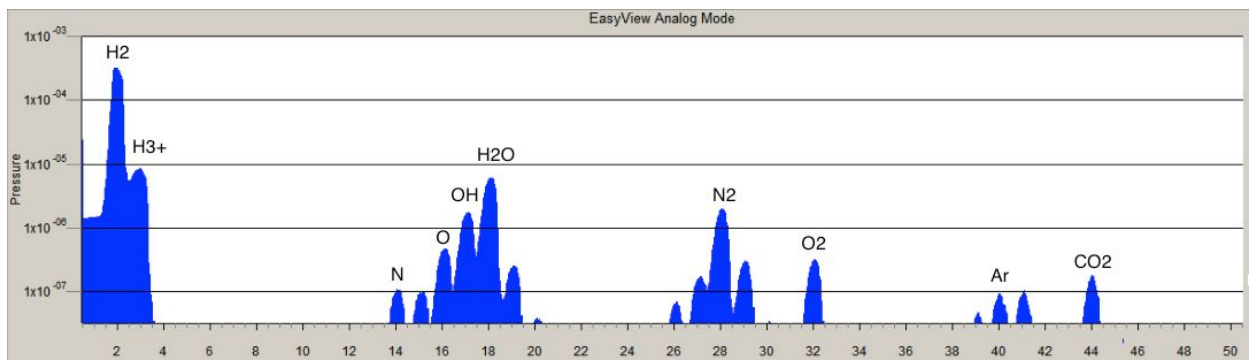

Fig. 11. Typical RGA mass spectrum of gases pumped out from the X-ray tube. Hydrogen (m=2 and m=3 $H_3^+$ ion peaks) is the main component of the residual atmosphere. Water (m=18 cluster), nitrogen (m=28) and oxygen (m=32) are also present.

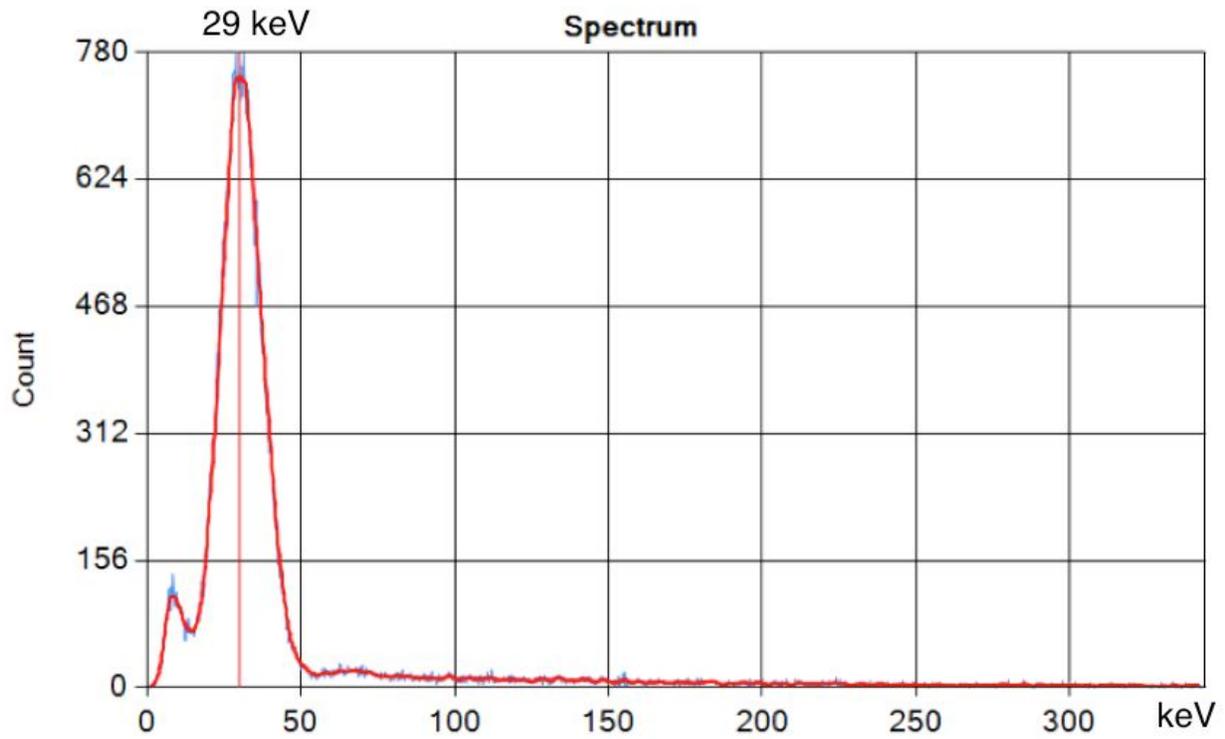

Fig. 12. Typical gamma spectrum showing 29 keV peak.

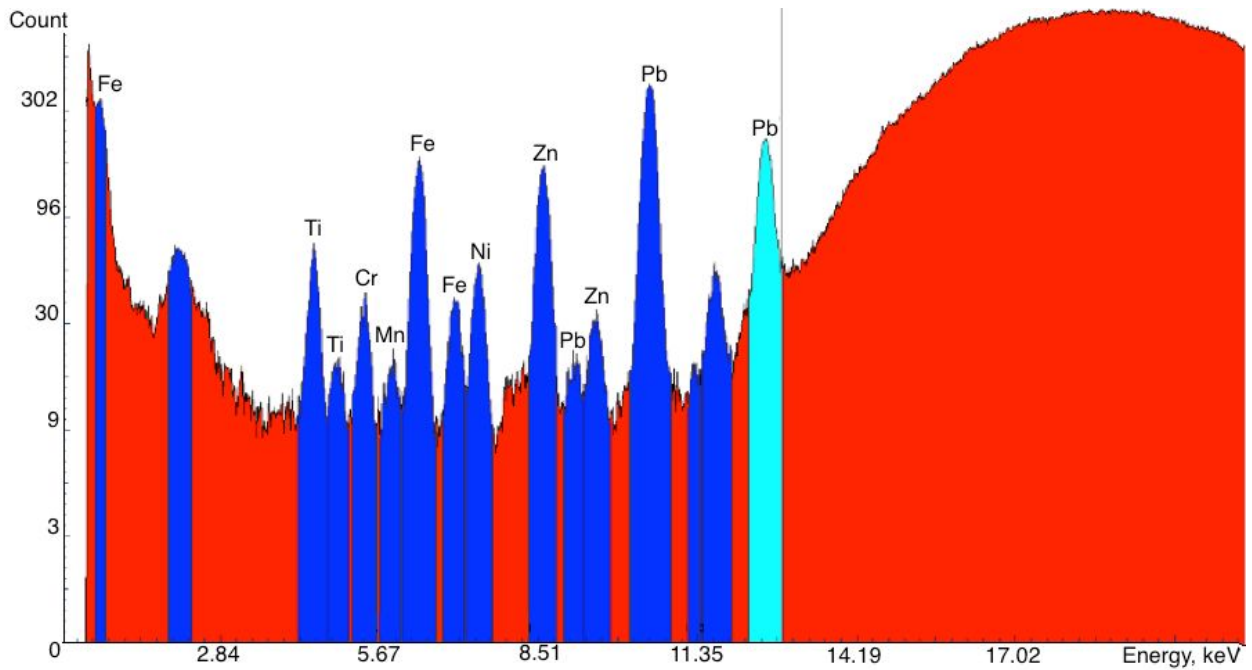

Fig. 13. Typical X-ray spectrum. Peaks originate from X-ray excitation of metals present in the anode, aonde enclosure, lead shielding and other materials of the tube.

**DISCUSSION**

Experimental results summarized in Table 1 indicate occasional presence of weak yet statistically significant neutron flux when a 20-30 kV / 20-30 mA discharge was sustained in our cold-cathode X-ray tube. The flux was not evident in all experiments and was not always repeatable. So far we were not able to determine which variables were important for causation of the effect, although it seems that that the material of the anode was important (i.e. we never detected the flux when copper anode was used).

In experiment conducted on 5/1/2018 using AlN ceramic we measured statistically significant ($P=0.024$) elevated neutron counts when the chamber was evacuated and filled with hydrogen at 0.1 Torr although there was no discharge. Increasing hydrogen pressure to atmospheric returned neutron counts to background. We recorded 10 sequences where we varied pressure in such way (0.1 torr for 'experiment' and 1 atm for 'background') and the P-value kept improving with each new sequence added to the sample set. In other words, hydrogen pressure inside the tube appeared as a trigger for the elevated neutron count. Unfortunately, this curious effect disappeared after we subjected the ceramic to 30 kV / 10-15 mA electron discharge and were not able to reproduce it with a new AlN sample.

Another curious observation was in elevated counts during the first minute of counting that was frequently associated with many experiment sequences. Some such counts results in multiple neutron events - Fig. 14 - that were rarely observed. However, despite spending significant effort on trying to make this a repeatable effect (e.g. by installing new anode and hoping that the first experimental sequence on the new anode will consistently produce elevated counts) we failed to achieve a statistically significant result. Also, long background counts produced occasional sequences with highly elevated counts that also contained multiple events similar to those reported on Fig. 14. Thus the most logical conclusion is that the first-minute elevated counts were mere coincidences associated with strong variations in background.

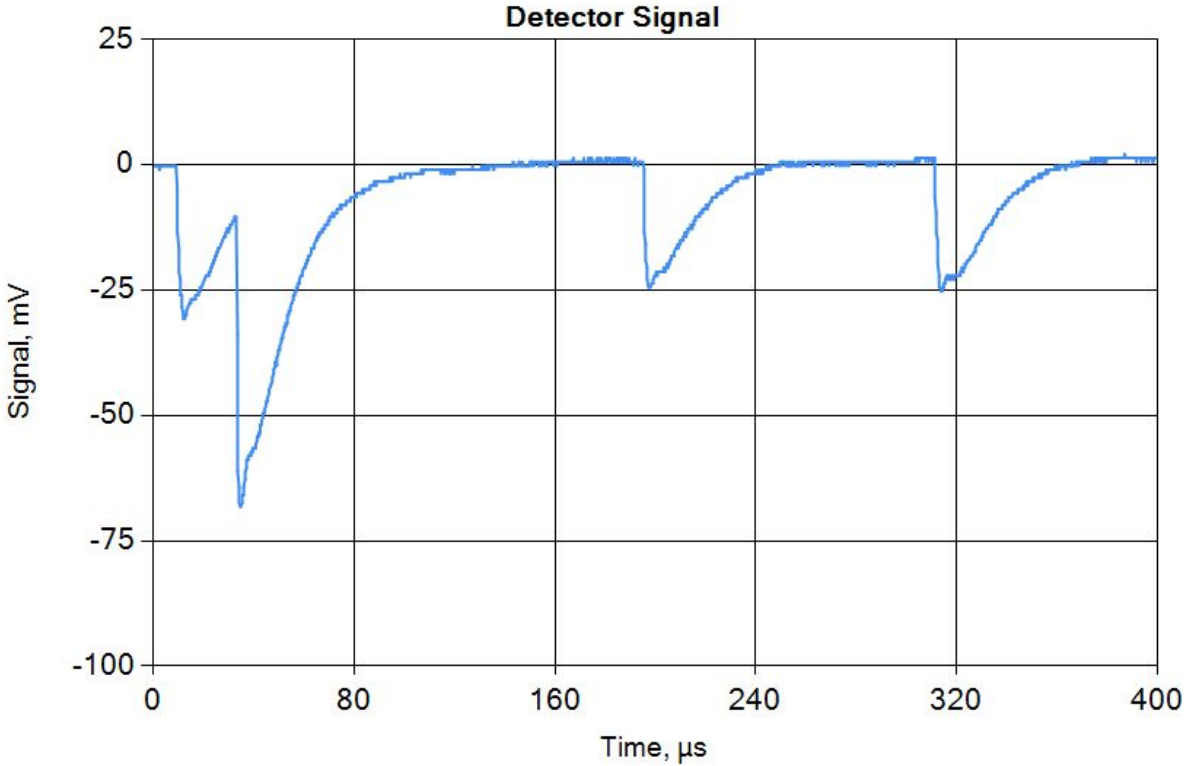

Fig. 14. Multiple neutron events associated with first-minute counts using titanium anode.

To rule-out very low-energy neutron flux that we might have rejected as x-rays (or lost due to absorption in lead and HDPE moderator) we placed a pair of BD-PND and a pair of BDT detectors right next to the tube thus obtaining a good solid angle coverage. These detectors are greatly superior to CR-39 in terms of sensitivity and ease of use and are entirely insensitive to electromagnetic interference [12]. However, these detectors did not develop any bubbles during the course of the experiments we conducted. Additionally in many experiments we have been deploying 1-mm thick 4x4" indium foil that we counted after 40 minutes of exposure using 3" NaI(Tl) gamma detector but failed detecting any above-background gamma counts indicative of neutron activation. Therefore we can reasonably conclude that we did not miss a very low-energy neutron flux that could have been absorbed by lead and HDPE moderator.

Lastly, to rule-out formation of ultra-low momentum neutrons that might have been absorbed directly by the anode we examined the anodes under electron microscope using the EDS analysis to look for elemental composition changes. On two occasions we have detected unusual elements present in the anode pit (i.e. in the crater carved by the electron beam striking the anode). In the case of nickel-alloy anode we observed huge quantities of rare earth (such as lanthanum and cerium) coating the anode pit bottom - Fig. 15.

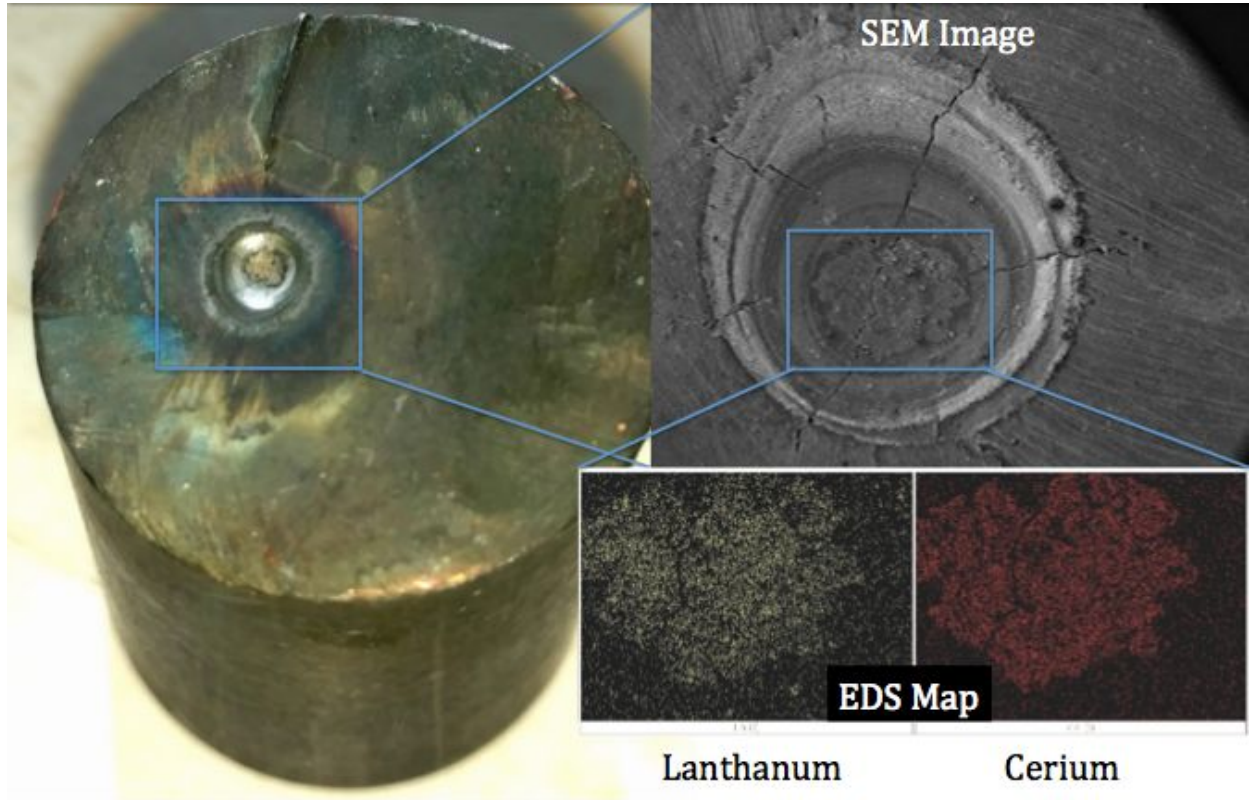

Fig. 15. Left: optical image of the nickel-alloy anode; top right: SEM image of the anode pit; bottom right: EDS maps of the anode pit showing lanthanum (yellow) and cerium (red) content.

When we examined the stainless steel anode we have discovered two iron droplets that looked like they were expelled from the anode pit - Fig. 16. We have analyzed these droplets using EDS and discovered that they are significantly enriched in thorium - Fig. 16. Closer examination revealed well localized areas on the droplet where thorium content exceeded 40% - Fig. 17.

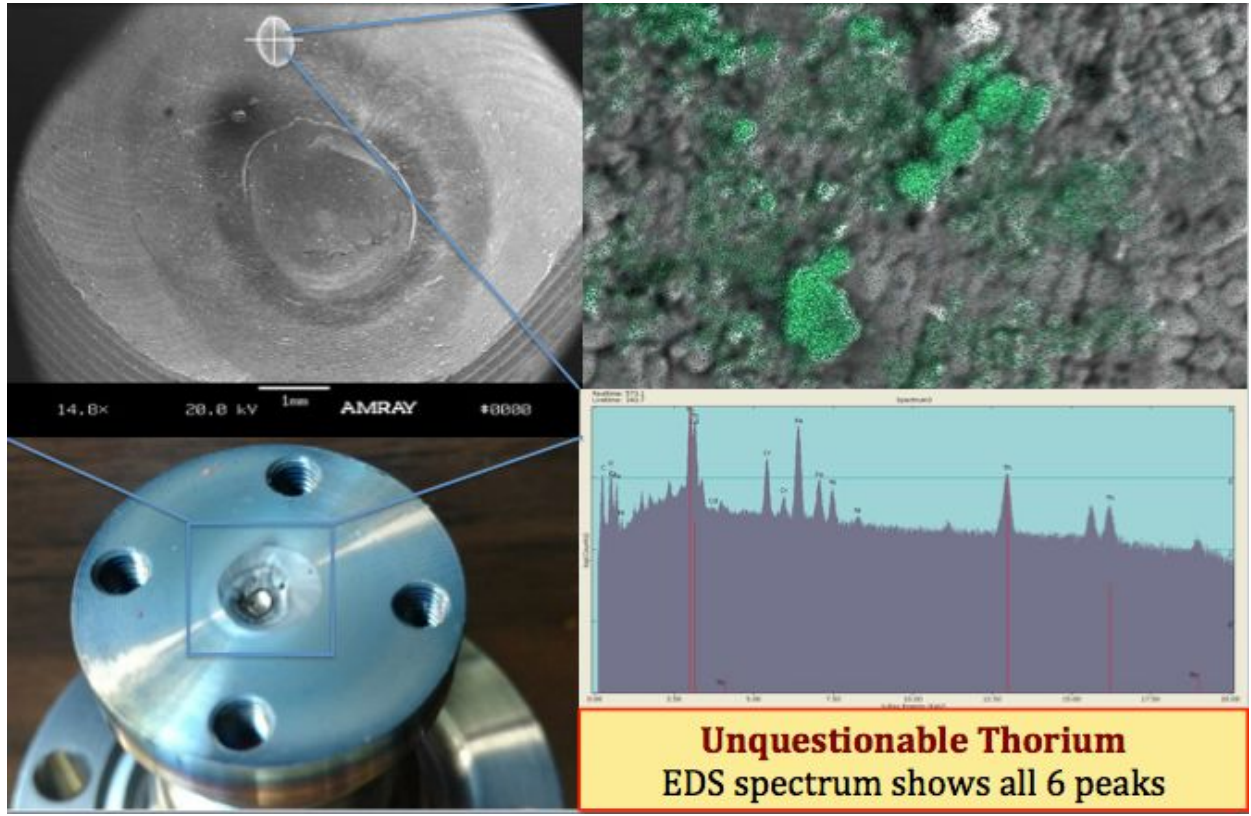

Fig. 16. Bottom left: optical image of the stainless anode; top left: SEM image of the anode pit featuring the iron droplet just above the pit (bright oval) ; top right: magnified SEM image with overlaid EDS map (green) showing relative thorium content.

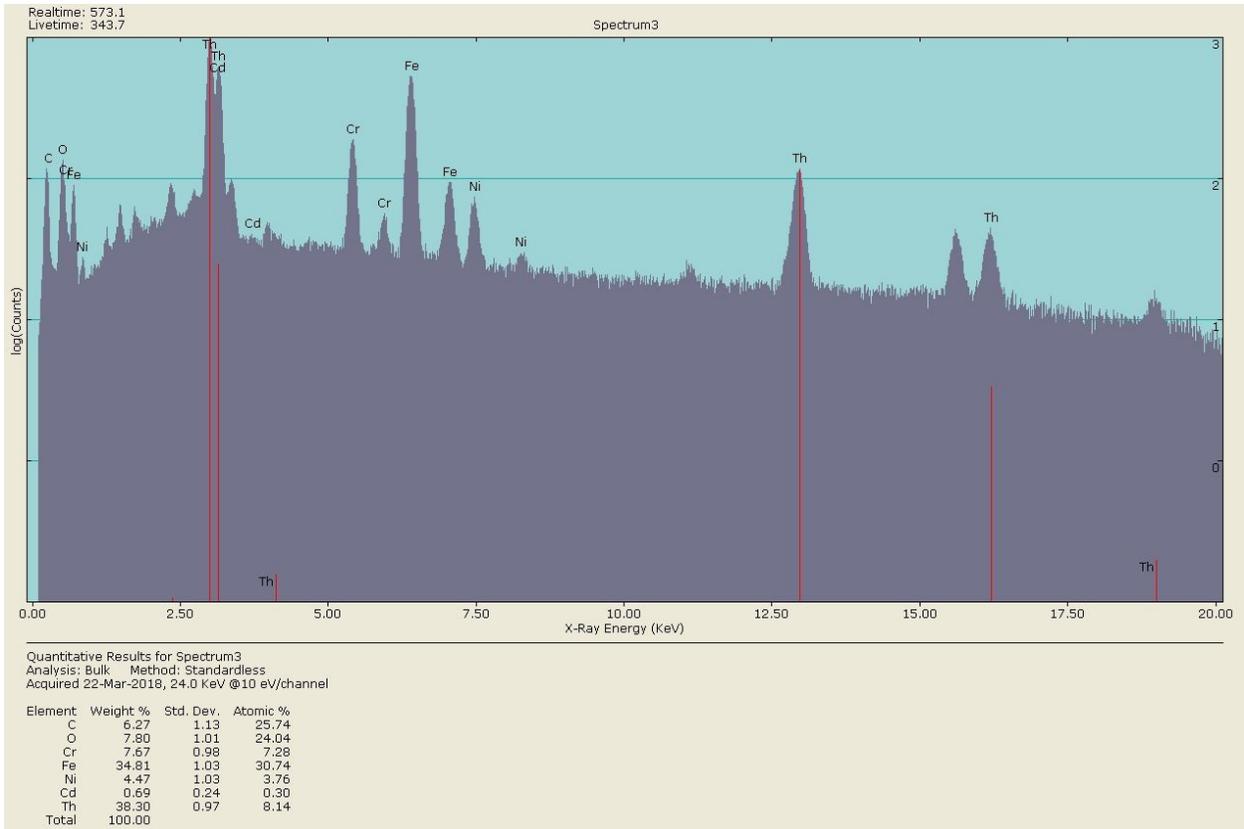

Fig. 17. EDS spectrum of iron droplet surface showing significant thorium content; all 6 thorium EDS peaks are clearly resolved.

The presence of thorium on the stainless anode is inexplicable from any plausible theoretical point of view. Therefore the only logical conclusion that we can draw that it must have been a contamination that was present on the stainless anode surface prior to the experiment. After all we did have pure thorium wire and thoriated tungsten in our lab.

The case of lanthanum and cerium on nickel-alloy anode proved more interesting but also explainable. Although the standard 20Cb3 alloy specification lists nickel, chromium, iron, copper, neodymium, molybdenum, manganese and silicon as major components (>1% composition) we discovered that some manufacturers add small (<1%) quantities of rare earth to improve the alloy characteristics. This admixture of rare earth is small enough not to be distinguishable by EDS when one analyzes bulk sample. However, when the electron beam melts and evaporates metals at the bottom of the anode pit, the metals with lower boiling temperature evaporate first leaving much harder to melt rare earths in place. Thus, the electron beam performs an effective enrichment of the alloy due to fractionation by boiling temperature. Hence the bottom of the anode crater ended up being covered with rare earth crust.

**ERNEST STERNGLASS EXPERIMENT: REPLICATION AND RE-EXAMINATION**

It is worth mentioning that our experimental setup was virtually identical to the one used by Ernest Sternglass at Cornell [14]. Following Rutherford, Sternglass believed neutron to be a bound state of a proton and an electron and therefore was hoping to prove this theory by synthesizing neutrons via $p^+ + e^- \rightarrow n^0$ reaction in a hydrogen-filled cold-cathode X-ray tube. Although Sternglass never published his results one can glean enough detail from his notebooks [15].

Because Sternglass shared Einstein's dissatisfaction with quantum theory Louis De Broglie invited him to work at his institute on semiclassical approach. Building on Bohr-Sommerfeld theory Sternglass developed a purely electromagnetic model of $\pi^0$-meson that accurately predicted its spin and mass [16]. Later Sternglass has published equally successful models for $\mu$-meson, *K*-meson, $\eta$-meson and their resonances [17].

Sternglass operated a very similar cold-cathode X-ray tube powered by an AC coil. To detect neutrons he placed silver and indium foils next to the tube and then counted the foils with a Victoreen IB85 geiger counter to look for neutron activation. Above background counts decaying to baseline with the rate consistent with half-life of the activated material (~3 minutes for silver, ~54 minutes for indium) would indicate presence of neutron flux.

Although Sternglass never developed a similar model for neutron he viewed his experiments at Cornell as a success [14] and claimed to have registered above-background neutron activity on the order of 1 CPM [15]. Intrigued by Sternglass' claims we have re-examined the raw data preserved in his notebooks but found no evidence supporting the statistically-significant above-background signal. Small sample sizes, selection bias and total lack of significance analysis have lead Sternglass to wrong conclusions about the meaning of his counts. On Fig. 18 we present Sternglass' processed plot of neutron counts and the corresponding raw counts. The plot does not show a clearly recognizable exponential decay and when we subject the activated silver and the background counts to Student's T-test we obtain a P-value of 0.23.

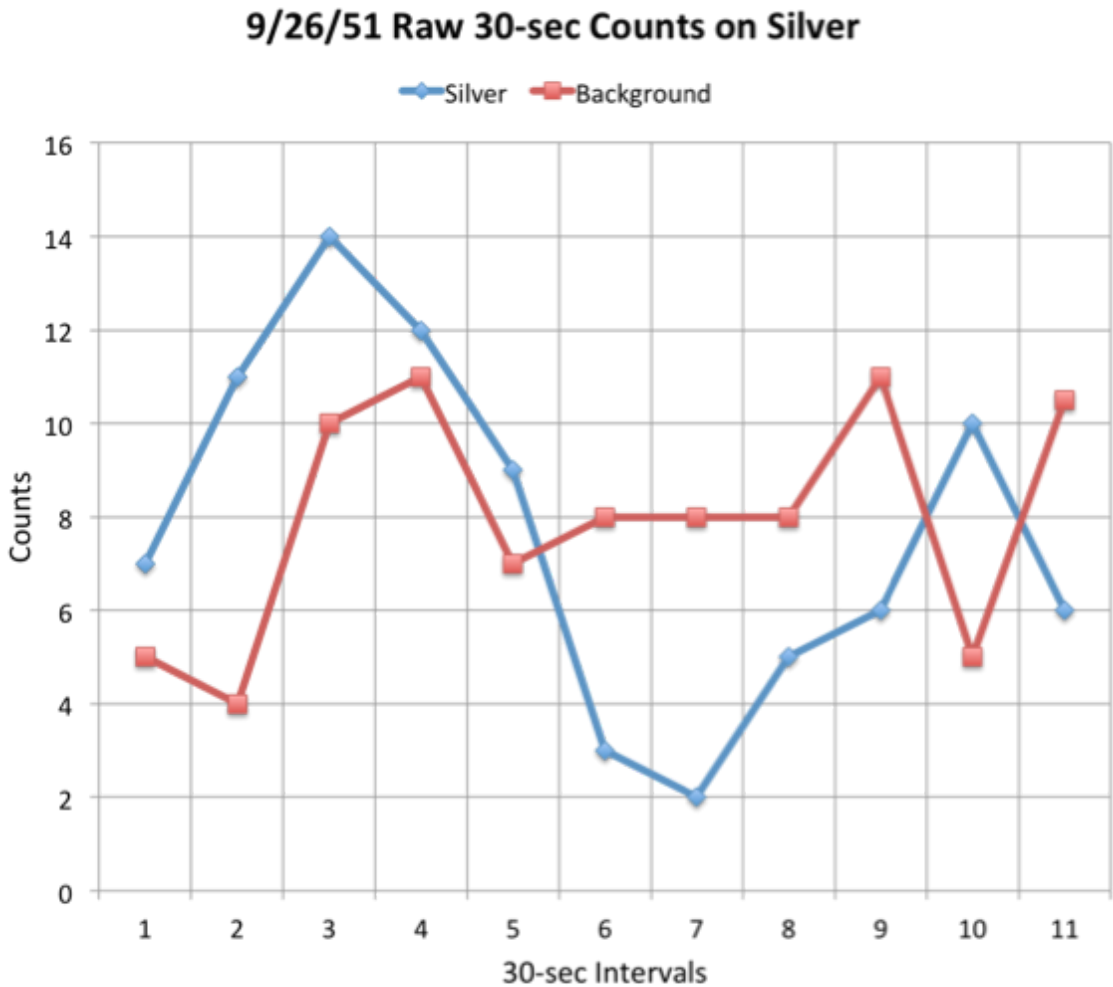

Fig. 18. Sternglass' processed data vs. raw counts.

For the sake of completeness we decided to reproduce Sternglass' neutron counting methodology exactly by wrapping silver foil exposed to the hypothesized neutron flux around Victoreen IB85 geiger counter and by pure chance arrived at plot that looked remarkably similarly to Sternglass' - Fig. 19. Subsequent tests yielded different count patterns, some consistent and some not with the decay curve characteristic of activated silver (which is not surprising considering the small sample size). P-value remained close to 0.5 and did not improve with the increase of the sample size.

However, it took a control experiment to completely discredit Sternglass' measurement methodology: we were unable to register any above-background counts from silver foil activated using 1 mCi Po-Be source neither by using the Victoreen geiger counter (which is rather insensitive) nor by employing a much-more sensitive Bicron NaI(Tl) 3" gamma probe. This is not surprising since compared to indium-115, helium-3 or boron-10, silver-107 and

silver-109 have rather small neutron capture cross sections. Therefore much larger flux (e.g. equivalent to 1 Ci Pu-Be source) is necessary to induce activity detectable with a geiger counter.

Based on the evidence presented in this section we find no support for neutron formation in Sternglass' original experiment.

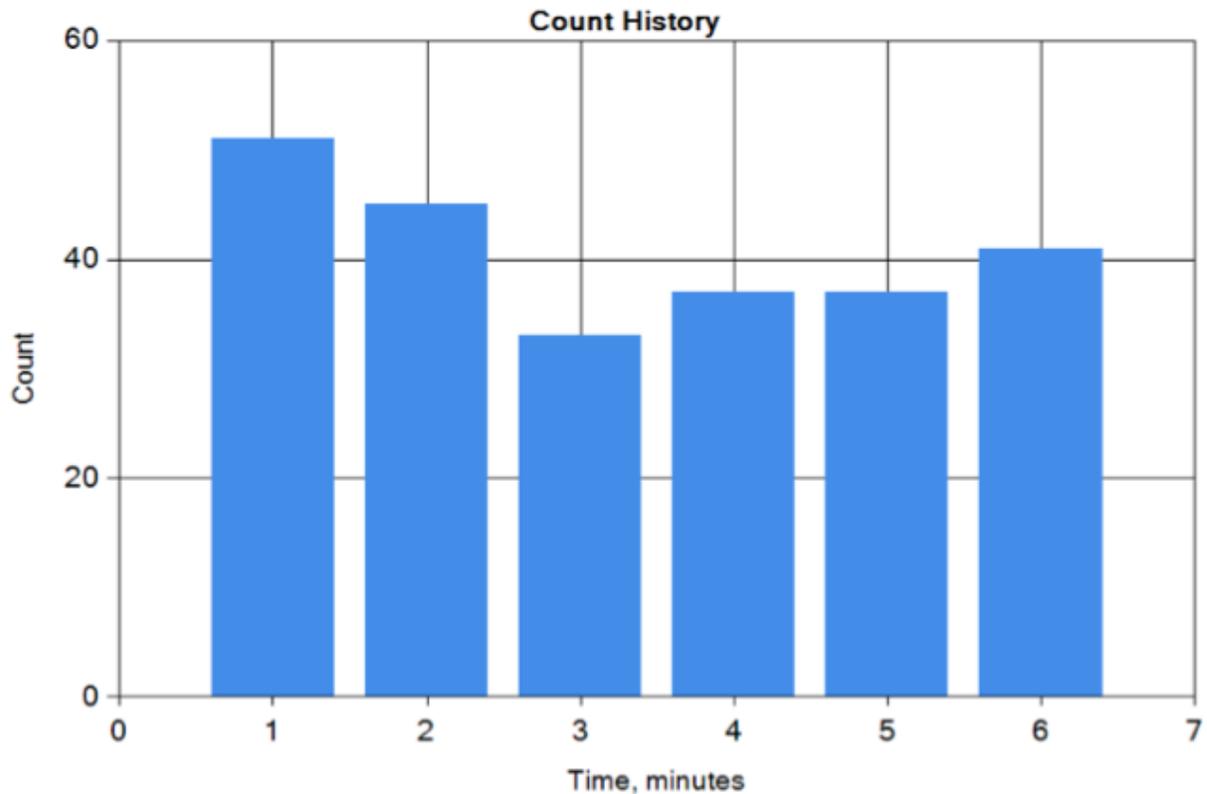

Fig. 19. Counts obtained by following Sternglass' methodology.

**CONCLUSIONS AND FUTURE WORK**

We have proposed a solution to neutron formation puzzle in atmospheric discharge when the discharge voltage is below the threshold of photonuclear reactions. We point out that rapidly changing magnetic vector potential creates strong electrokinetic field that can accelerate free electrons to energies beyond the threshold of photonuclear reactions (i.e. over 10.5 MeV).

In order to detect the purported neutron flux we have developed highly sophisticated and extremely sensitive neutron detector system that can reliably and in real-time identify neutron fluxes smaller than 0.1 CPS above background. This system was designed to be impervious to EM noise and therefore is suitable for use in high-noise environments (such as high voltage / high current discharges) where traditional instrumentation cannot be trusted.

We have applied our neutron detector system to measure neutron emission from 20-30 kV / 20-30 mA discharge sustained in dilute atmosphere and indeed registered weak but statistically significant above-background neutron flux in several experiments.

Assuming that our theoretical analysis is correct the next logical step in this research is to use our equipment to measure neutron flux from high-current atmospheric discharge (e.g. 20-100 kV / 10-30 kA). We expect the magnitude of the flux to go up with the increase in current and decrease in stability of the discharge (i.e. under the conditions that maximize $\partial A/\partial t$).